\documentclass[12pt]{article}
\usepackage{amssymb,amsmath}
\usepackage{euscript}

\begin{document}

\title{Reference frames in classical and relativistic physics}

\author{{\bf O.I. Drivotin}\\
{\small Faculty of Applied Mathematics and Control Processes, St.-Petersburg State University,}\\
{\small  35, Universitetskii pr., Petergof, St.-Petersburg, 198504, Russia}\\
{\it\small  e-mail: drivotin@ya.ru}
}

\maketitle
\begin{abstract}
Formal definition of the reference frame is given.
This definition is valid for nonrelativistic and relativistic cases.
Proposed definition allows using wide classes of reference frames
without restriction to inertial, uniformly accelerated or rotating frames.
A conception of the system of coordinates associated with a reference frame is introduced.
It is shown that one of these coordinates can be regarded as temporal coordinate,
and the others as spatial ones.
It is demonstrated that 
in relativistic case nondiagonal spatial-temporal components of the metric tensor
are always equal to zero. 
Inertial, accelerated, and rotating reference frames are considered
as examples.

\end{abstract}

%\keywords{Reference Frame; Metric Tensor; Einsten's Equations.}
%\pacs {04.20.-q, 04.20.Cv}
%\submitto{\EJP}
 
%\maketitle
 
\section{ Introduction }

The term "reference frame" is widely used in physics.
%but it has not rigorous definition.
%The reference frame is commonly introduced as an observer,
%which can measure time intervals and distances with clocks and rules.
%Such approach cannot be adopted as mathematically correct,
%because it does not contain precise definitions of the measured values.
%
In works \cite{math,vlad} the reference frame is associated with a congruence of observers
in the spacetime.
%(see, for example, \cite{math,vlad}).
In the present work such approach is used and developed.
It is shown how to set correspondence
between event and its four coordinates
such that one of them is temporal coordinate and the others are spatial ones
in the given reference frame.
Proposed definition of the reference frame
relates equally to nonrelativistic and relativistic cases.
 
Another problem concerned in this paper is the physical sense 
of the metric tensor in relativity theory. 
%The metric tensor is the main object in relativity theory.
The metric tensor is usually considered as tensor with given components,
for example, as a solution to the Einstein equations.
%At such approach, physical sense of the metric tensor remains unclear.
%But following question is unanswered:
%how should one specify the components of the metric tensor
%in a reference frame?  
In Einstein's work \cite{ein1}, it is stated that 
"in the general theory of relativity, space and time cannot be defined
in such a way that differences of the spatial coordinates
can be directly measured by the unit measuring rod,
or differences in the time coordinate by a standard clock."
By this reason, physical sense of the metric tensor components
can be determined only locally, and remains unclear in extensive region.
In the present work the contrary approach is proposed.
It is explained how to specify
components of the metric tensor in the coordinate system associated with a reference frame.
Components of the metric tensor in this system  have plain physical sense.
It is turned out also that 
nondiagonal spatial-temporal components of the metric tensor
are always equal to zero.

The article is organized as follows.
In the next section,
definition of the reference frame is given.
The definition includes such concepts as main observer
and congruence of local observers.
The system of coordinates associated with the reference frame,
and the configuration space associated with the reference frame
are defined.

In the following two sections,
reference frames
in classical and relativistic theories
are considered separately
(classical here and further means nonrelativistic).
The difference between these two cases
arises from peculiarities of the theories,
particularly from relativity of simultaneity in the theory of relativity.
%and velocity restriction in relativistic theory.
 
Three examples are considered.
They are the inertial frame in the third section,
the accelerated and rotating frames in the fifth section.
These examples should be regarded not only as illustrations,
but as important conceptions consistent with the notion of the reference frame.

Discussion and comparison 
with other approaches are given in the last section.

\section{ General Definition of the Reference Frame }

In modern physical theory 
the spacetime is considered as differentiable manifold
points of which are events.
Some set $M$ being manifold means that for each point $x\in M$
there exist a set $D$ containing $x,$
an open domain $U$ in $R^n,$ and one-to-one mapping 
$f: D\mapsto U\subset R^n.$
The set $D$ can be regarded as neighborhood of the point $x.$
Therefore, the mappings $f$ induce a topology on $M.$

The mapping $f$ including the domain $D$ is called
a system of coordinates,
and components of $f(x)$ as a vector in $R^n$
are called coordinates of the point $x$ 
in this system of coordinates.
Denote coordinates of a point $x$ by $x^i,$
$i=\overline{1,n}.$
For the four-dimensional spacetime, we shall use traditional numeration: $i=\overline{0,3}.$

Differentiable manifold is a manifold
such that for each pair of mappings
$f_{\alpha}: D_{\alpha}\subset M \mapsto U_{\alpha}\subset R^n$
and $f_{\beta}: D_{\beta}\subset M \mapsto U_{\beta}\subset R^n$
functions of transition from any system of coordinates to another coordinates
$f_{\alpha}^{-1}\circ f_{\beta}$
defined on the set $f_{\beta}^{-1}(D_{\alpha}\cap D_{\beta})$
are continuously differentiable.

Thus, every system of coordinates
is admitted for a description of the spacetime
if the transition mappings
between it and another admissible systems of coordinates
are continuously differentiable.

Nevertheless, there exist
systems of coordinates
constructed on the basis of
some physical considerations.
They can be chosen as
some initial systems of coordinates,
in which components of tensors representing physical values
can be specified,
and from which one can
pass to another systems using
coordinate transformation.
Such systems of coordinates are used
in the traditional description of the spacetime,
when one of the coordinates is the time
and the others are regarded as spatial coordinates.

Time coordinate in relativity theory 
and spatial coordinates are determined depending on a reference frame.
Therefore, a reference frame is a way to choose such
system of coordinates that one is temporal and another is spatial. 
%Another words,
%a reference frame is a mapping of some domain in the spacetime  $D$ 
%to $(t_0,T) D_0$
%where $(t_0,T)$ is real interval and $D_0\subset R^3.$
%Using of a reference frame
%significantly simplifies description
%of the physical processes,
%espessially in nonrelativistic theory.

In what follows, we shall concern ourselves how to construct such mapping.
Consider an observer,
by which we mean point particle carrying clock,
and call this observer the main observer.

Assume that there exists such (open) domain $D$
of the spacetime that for each event 
in this domain one can find simultaneous event
on the worldline of the main observer.
The contrary assertion means
impossibility of time measuring outside worldline of the main observer.
This assumption has fundamental character.
It is always held in classical theory.
In relativity theory, it may be supposed
that this assumption is valid not always.
Nevertheless, in some cases it is valid.
At least, it is valid for the flat spacetime.
As we shall seen later, it is valid for many other cases,
for example, for the spacetime with the Schwarzschild metrics.

All simultaneous events form some set.
Let's call this set the space of simultaneous events,
or the temporal layer, and denote it by ${\EuScript T}_t,$
where $t$ is time of the main observer for this layer.
A temporal layer is defined for each event in the domain $D.$
Therefore, the temporal layers form a foliation of the domain $D:$
$$
%\equation
D=\bigcup\limits_t{\EuScript T}_t.
%\endequation
$$

Assume that each temporal layer is connected differentiable
3-dimensional manifold,
%Therefore, 
and it is possible to introduce three coordinates
on each temporal layer.
 
Consider a set of observers.
If for each point $x\in D$ 
there exists unique observer
whose worldline passes through this point,
it is said that the observers form a congruence of observers.
It means that the domain $D$ can be represented 
as union of nonintersecting worldlines of observers.
Another words, worldlines of the observers fill up the domain $D.$
One of these observers is the main observer,
the others will be called local observers.

Assume that there exists congruence $\EuScript C$
of observers, one of which is the main observer,
such that all events on each temporal layer are also simultaneous
for each local observers.
A main observer, a set of temporal layers corresponding to the main observer,
and a congruence of observers
satisfying formulated here condition will be called a reference frame.

Consider a reference frame
and choose some temporal layer ${\EuScript T}_0.$
Call chosen layer the configuration space
associated with given reference frame.
Introduce system of mappings ${\EuScript F}_t$
of temporal layers ${\EuScript T}_t$
to the layer ${\EuScript T}_0$
defined as follows:
we assign to each point $x\in {\EuScript T}_t$
a point ${\EuScript F}_t(x)$
that is intersection of the wordline of the observer from $\EuScript C$
passing through $x$ with ${\EuScript T}_0.$
Time $t$ measured by the main observer 
and coordinates in the configuration space constitute
a system of coordinates
that we shall call the system of coordinates
associated with the reference frame.

If a reference frame is specified,
time $t$ measured by the main observer can be regarded as temporal coordinate,
and coordinates in the configuration space can be regarded as spatial ones.
What does it mean?
The local observers can regard
physical processes in the spacetime
described in these coordinates  
as spatio-temporal,
because for the local observers temporal layers
of the main observer are also layers of simultaneous events.
Thus, these coordinates are temporal one and spatial ones in local sense.

The main observer can describe physical processes
using local spatial-temporal descriptions of all local observers.
It is apparent in view of the fact that 
differential equations of mathematical physics
have tensor form and, therefore, are local equations.
Such global description can be called also spatial-temporal one.
In this sense these coordinates are temporal and spatial ones.

Such approach is traditional in physics.
Many physical values are defined as tensors in the configuration space.
Components of tensors are defined locally.
Specifying tensor components on the base of a physical sense
requires that one coordinate be temporal and the others be spatial 
from the point of view of the local observer.
For example, components of the tensor of electromagnetic field $F$
are components of the electric field ($F_{0i},$ $i=1,2,3$)
and components of the magnetic flux density ($F_{ik},$ $i,k=1,2,3$),
only if coordinate $x^0$ is temporal one and coordinates $x^i,$ $i=1,2,3$ are spatial ones in local sense.
If it is not so, components $F_{ik},$ $i,k=1,2,3,$ 
will be mixture of components of the electric and magnetic field.

In addition, the distance can be measured by a local observer
only between two close simultaneous events.
Therefore, distance measuring on the temporal layer is not possible,
if assumed condition of simultaneity is not valid.

If a reference frame is specified,
physical processes can be studied in the configuration space.
One can watch how does a mathematical object $A$ describing a process
change depending on $t.$
For example, 
the motion of a particle
can be studied
examining images of positions
of the body on different temporal layers $\overline x={\EuScript F}_t(x).$
Then set of images corresponding to different $t$
represents the trajectory of motion in the configuration space,
and the time $t$ measured by the main observer
can be taken as a parameter of the trajectory.
Such mathematical model of the motion
is widely used in physics for a long time.
When this model is used, parameter $t$ is regarded as time,
and coordinates in the configuration space are regarded as spatial ones.
It is fully consistent with our approach.

Choose a reference frame and 
assume that at each instant each observer can measure distances
from him to all close points lying on the corresponding layer.
A criterion of closeness will be formulated further.

Distances between close points can be described
by the 3-dimensional metric tensor defined 
on the temporal layer as on 3-dimensional manifold.
Let us give formal definition
of bilinear form, which we shall call the metric tensor.
Consider a point $x$ and a point $\xi$ that is close to $x;$
$x,\xi\in M,$ $dim M=n.$
Assume that we can connect these points by a smooth line $y(\lambda):$
$x=y(\lambda_0),$ $\xi=y(\lambda_0+\delta\lambda).$
Criterion of smoothness will be given also further.
Let introduce  vector $\delta x$ of displacement
of the point $\xi$  relative to the point $x:$
$\delta x=v\delta\lambda\in T_xM$
where $v=dy/d\lambda$ is the tangent vector to the line at the point $x.$
Here $T_xM$ denotes the tangent space of the manifold $M$ at the point $x.$

Nondegenerate symmetric bilinear form $g(u,v),$ $u,v\in T_xM,$
such that for any point $\xi$ close to the point $x$
\equation
g(\delta x,\delta x)=\delta r^2
\label{eq:defmetric}
\endequation
with an accuracy up to terms of higher order $o(\delta r^2)$
will be called the metric tensor.
Here $\delta r$ is the distance between  $\xi$ and  $x.$

Assume that there exists such system of coordinates
that deviations of components of the metric tensor
from corresponding elements of the matrix
$$
\|g_{ik}\|={diag}(1,\ldots,1)
$$ 
are small compared to $1.$
Let us call such coordinates locally Cartesian coordinates in the domain $D.$
A line that can be represented in locally Cartesian coordinates in the form
$x^i=x_0^i+\lambda v^i,$ $x\in D$  where $v$ is some vector
will be called a smooth line.
Each pair of points in $D$ that can be connected by a smooth line will be called
close each to other.

When components of $g$ in some coordinates changes slowly
along the smooth line connecting the points,
equality (\ref{eq:defmetric}) can be written in the form
$$
\delta r^2=\sum_{i=1}^n\sum_{k=1}^ng_{ik}\delta x^i\delta x^k.
$$

\section{Reference Frames in Classical Physics}

One of the basic postulates of nonrelativistic theory
is absoluteness of the simultaneity.
It means that 
if one observer regards two events as simultaneous,
another observer who is able to measure time of these events
also regard the events as simultaneous.
It is usually assumed that
every observer can measure the time of all events,
and that rate of clocks of all observers is the same,
but the contrary assumptions are also acceptable.

Then the spacetime can be represented as union of the temporal layers
${\EuScript T}_t$ \cite{wang}:
\equation
{\EuScript R}=\bigcup\limits_t{\EuScript T}_t,
\label{eq:classicspace}
\endequation
where $t$ is the time for this layer.

Thus each congruence of observers 
specifies a reference frame. 
One of these observers can be taken as the main observer, 
and the others as local ones.

The time $t$ can be regarded
as temporal coordinate  
for any reference frame.
But choice of spatial coordinates depends on a reference frame.
For example, spatial coordinates of main and local observers
participating in constructing of a reference frame 
don't change in system of coordinates associated with this reference frame
and can change in coordinates associated with another reference frame.

As it was assumed earlier, 
on each temporal layer ${\EuScript T}_t$
 one can introduce metrics,
that is the nondegenerate symmetric tensor
$g_{ik},$ $i,k=1,2,3,$
specifying distances between simultaneous events.

As an important example,
consider the inertial frame.
Let the domain $D$
%in which coordinates are defined
is such that for layer ${\EuScript T}_0$ all points are close 
to the position of the main observer on this layer.
%Take locally Cartesian coordinates of ${\EuScript F}_t(x)\in {\EuScript T}_0$
%as spatial coordinates of an event $x\in D.$ 

Let us call a reference frame
such that velocity in the configuration space $v$ of every isolated particle doesn't change
the locally inertial frame.
It means that Cartesian components of the velosity conserve,
because Lagrangian of an isolated particle with mass $m$
in such frame has the form $L=mg_{ik}v^iv^k/2$
(summation is meant on coincident indices).
Inertial frame is the particular case when the domain $D$ is unbounded.

The difference between four-dimensional velocity vector
of some point particle  $u=dx/dt\in T_x{\EuScript R}$
at the point $x$
and four-dimensional velocity vector of the local observer 
$u_O(x)\in T_x{\EuScript R}$ at this point has only spatial components.
So it can be considered as three-dimensional vector
$u-u_O(x) \in T_x({\EuScript T}_t).$
It is easy to understand that this vector is $v.$

Take two inertial frames and associated systems of coordinates
with Cartesian spatial coordinates.
Four-dimensional velosity of a particle is 
$$
u=(1,v^1,v^2,v^3)^T=(1,\tilde v^1,\tilde v^2,\tilde v^3)^T
$$
(index $T$ here and further denotes transposition, tilde indicates the second reference frame).
Here $v^i={\rm const}, \tilde v^i={\rm const},$ $i=\overline{1,3},$
according to definition of the inertial frame.

Consider a particle at a point $x\in D.$
We have $u-u_{\tilde O}=u-u_O-(u_{\tilde O}-u_O).$
Therefore $(u_{\tilde O}-u_O)=V,$
where $V^i=v^i-\tilde v^i={\rm const},$ $i=\overline{1,3}.$
It may be interpreted as uniform motion
of the second frame relative to initial frame.

Applying the contravariant law of transformation of vector components
we get following equation for coordinate transformation functions:
\equation
\tilde u^i=\frac{\partial \tilde x^i}{\partial x^j}u^j=v^i-V^i,\quad
i=\overline{1,3}.
\label{eq:transin}
\endequation
The unique solution to (\ref{eq:transin})
is:   $\tilde x^i=x^i-V^i x^0,$ $i=\overline{1,3}.$
This law is known as Galilean transformation.

\section{Reference Frames in Relativity Theory}

In relativity theory, two approaches are possible.
According to the standard approach,
it is assumed that there exists a four-dimensional metric tensor $g$
defined at each point of the spacetime
such that matrix of its components in some coordinates
has the form \equation
\|g_{ik}\|={\rm diag}(1,-1,-1,-1),\quad i,k=\overline{0,3}
\label{eq:psevdodiag}
\endequation
at this point.
In this case, one can do without reference frames,
because all tensor equations are valid being
written componentwise in every system of coordinates.
But the sense of components
of the  metric tensor and another tensors doesn't have simple interpretation.

We shall follow to the second approach,
which consists in direct definition of the components of the metric tensor
in coordinates that have plain physical sense.

First of all, let construct a reference frame
in some domain $D$ of the spacetime,
as it is described in the section 2.
As a result, we get a representation of the domain $D$ 
as a union of disjoint layers,
each of them representing three-dimensional surface
describing by the equation $t(x)={\rm const}.$
Here $t(x)$ is time measured by the clock of the main observer
for the layer on that the event $x$ lies. 
As distinct from the nonrelativistic case,
events on the temporal layers are not simultaneous for observers of another reference frames.

Though events on all temporal layers
are simultaneous for all local observers of this reference frame,
time intervals between layers can differ
from the points of view of different local observers.
Let introduce the coefficient $g_{00},$
which will characterize clock rate of a local observer
relative to  clock rate of the main observer:
$\delta \tau=\sqrt{g_{00}(x)}\delta t.$
Here $\delta t$ and $\delta \tau$
are time intervals between close layers
according to clocks of the main observer and a local observer correspondingly
(assume
 that $\delta t$ and $\delta \tau$ have the same sign).

Let's take the time $t$ as temporal coordinate of a point $x,$
with an accuracy up to multiplier $c,$
sense of which will be revealed further:
$x^0=ct.$
As before,
let's take coordinates of the image  of the point $x$
in configuration space as coordinates $x^i,$ $i=1,2,3.$
They are spatial coordinates.

As well as in nonrelativistic theory,
assume that 
the distance between a pair of close points is defined on each temporal layer,
and it can be expressed by the three-dimensional metric tensor,
which is defined on this layer:
$$
\delta r^2=\sum_{i=1}^3\sum_{k=1}^3g_{ik}\delta x^i\delta x^k
$$
with an accuracy up to terms of higher order
under formulated above condition of slow variation of $g_{ik}.$

Consider two close events on different temporal layers.
%What does the closeness of these events mean?
A criterion of closeness will be formulated further
in similar manner as for nonrelativistic case.
Let formulate some conditions of closeness in advance.
Take two observers whose worldlines $l_1$ and $l_2$ pass through the events.
Take temporal layers which contain the events, and denote one of them by ${\EuScript T}_1.$
Firstly, demand that points $l_1\cap {\EuScript T}_1$ and $l_2\cap {\EuScript T}_1$ 
be close one to other in the same sense as in nonrelativistic case.
%Secondly, demand that difference of the values of $g_{00}$ for the points
%is small compared to $g_{00}.$
As to the next condition, 
nonstrictly speaking,
it means that components $g_{00}$ and $g_{ik},$ $i=1,2,3$
vary slowly in some domain containing events closeness of which is examined,
particularly, along smooth line connecting
points  $l_1\cap {\EuScript T}_1$ and $l_2\cap {\EuScript T}_1.$
 
Let introduce the interval between
two close points as follows:
\equation
\delta s^2=|c^2\delta\tau^2-\delta r^2|=
|g_{00}(\delta x^0)^2-\sum_{i=1}^3\sum_{k=1}^3g_{ik}\delta x^i\delta x^k|
\label{eq:ds}
\endequation
with an accuracy up to terms of higher order.
One  of the basic principles of relativity theory
is that there exists such $c$ 
that interval between two close events is the physical value
which depends only on these events.
Interval invariance can be proved experimentally
by measuring time intervals and distances entering into expression (\ref{eq:ds})
in various reference frames.
The value of $c$ can be found from these measurements.
From the other hand,
the value of $c$ can be found 
from measurement of the light velocity,
as the electrodynamics equations contain the metric tensor.
The interval is called timelike, spacelike, or lightlike,
if the value $c^2\delta\tau^2-\delta r^2$
is positive, negative, or equal to $0$
correspondingly.

%Thus, we demonstrate how to introduce the interval in the spacetime.
Let us call the symmetric twice covariant tensor
satisfying the condition
$$
\delta s^2=\pm g(\delta x,\delta x)
$$
up to terms of higher order
the four-dimensional metric tensor.
Here $\delta x$ is the displacement vector of one point
relative to another close point.
Sign plus or minus are chosen for
timelike and spacelike intervals correspondingly.

If differences between components of the metric tensor
and corresponding components of the matrix (\ref{eq:psevdodiag})
are small compared to $1$ in coordinates associated with some reference frame,
call such reference frame and coordinates locally Lorentzian ones.
As previously,
a line that can be represented in locally Lorentzian coordinates
in the form $x^i=x_0^i+\lambda v^i,$ $x\in D$  where $v$ is some vector
will be called a smooth line.
Each pair of points in $D$ that can be connected by a smooth line will be called
close one to other.

%Further we shall consider  Lorentzian coordinate systems only
%with domains such that each pair of events can be connected by a smooth line.
%Then all events in a Lorentzian coordinates domain 
%can be regarded as close each to other,
If the domain $D$ is unbounded, the reference frame is called Lorentzian.
Spacetime where one can specify a Lorentzian frame is commonly called the Minkowskian spacetime.
 
According to the definitions of the interval and of the metric tensor,
components of the metric tensor with one zeroth index are equal to $0$
in coordinates among which one is temporal, and three ones are spatial:
\equation
g_{0i}=0,\quad i=1,2,3.
\label{eq:gi0}
\endequation
The component $g_{00}$
characterizes the rate of clock of a local observer
%which spatial coordinates don't change
as compared with the rate of the main observer clock.
Components of the metric tensor in other systems of coordinates
can be found by corresponding transformation.

As it is follows from the foregoing,
if one of the components $g_{0i},$ $i=1,2,3$
is not equal to $0,$
then one cannot regard $x^0$ as a temporal coordinate,
and the rest three coordinates as spatial ones.
In this case they are some combinations of temporal and spatial coordinates.

In various problems,
components of the metric tensor are found as solutions to the Einstein equations.
Is it possible to construct a reference frame always in these cases?
It can be done,
%Construction of a reference frame is possible,
if construction of spaces of simultaneous events
(temporal layers) is possible.
Write an equation of these three-dimensional surfaces in the form
$\tilde x^0(x^0,x^1,x^2,x^3)={\rm const}.$

Assume that
%the basis vector of coordinate $x^0$ is timelike, and
$g_{00}>0$ in the domain under consideration.
Assume also that  $\tilde x^0(x^0,x^1,x^2,x^3)$ is continuously differentiable on 
$x^i,$ $i=\overline{0,3},$ and that
${\partial x^0}/{\partial \tilde x^0}\neq 0$
when $x^1,x^2,x^3$ are fixed.
Then we can take $\tilde x^0,x^1,x^2,x^3$ as new coordinates.
%The value $\tilde x^0$ can be regarded as a temporal coordinate,
%and $x^1,x^2,x^3,$ $i=\overline{1,3}$ --- as spatial coordinates.
Matrix of coefficients of the covariant law of transformation
to these coordinates has the form
\equation
\frac{\partial x}{\partial  \tilde x}=
\begin{pmatrix}
{\partial x^0}/{\partial \tilde  x^0}&
{\partial x^0}/{\partial \tilde x^1} &
{\partial x^0}/{\partial \tilde x^2} &
{\partial x^0}/{\partial \tilde x^3} &
\cr
0&1&0&0\cr
0&0&1&0\cr
0&0&0&1\cr
\end {pmatrix}.
\label{eq:trans}
\endequation

It follows from (\ref{eq:trans}) that $\tilde g_{00}=g_{00}({\partial x^0}/{\partial \tilde x^0})^2>0.$

As sought surface is a surface of simultaneous events, we have
$\tilde g_{0i}=0$ for $i=1,2,3.$
Expressing these components with use of (\ref{eq:trans}),
we get equations
\equation
0=g_{00}\frac{\partial x^0}{\partial \tilde x^i}+g_{0i}\frac{\partial x^0}{\partial \tilde x^0},
\quad i=\overline{1,3}.
\label{eq:vectors}
\endequation
Inverting matrix (\ref{eq:trans}) and taking into account (\ref{eq:vectors}),  we obtain
\equation
\frac{\partial  \tilde x^0}{\partial x^0}=(\frac{\partial  x^0}{\partial \tilde x^0})^{-1},\qquad
\frac{\partial  \tilde x^0}{\partial x^i}=\frac{g_{0i}}{g_{00}}
 (\frac{\partial  x^0}{\partial \tilde x^0})^{-1},\quad
i=\overline{1,3}.
\endequation
So, differential form describing sought surface
can be written in the form
\equation
T=\lambda(x)(g_{00}dx^0+g_{01}dx^1+g_{02}dx^2+g_{03}dx^3),
\label{eq:TForm}
\endequation
where 
$$
\lambda(x)=\frac 1{g_{00}}(\frac{\partial  x^0}{\partial \tilde x^0})^{-1}.
$$
The form (\ref{eq:TForm}) has noncovariant view,
because describes the spaces of simultaneous events,
and the simultaneity is relative.

If the form is considered in the domain
that satisfies the conditions of the Poincare lemma,
it is necessary and sufficient for a possibility of its integration that
there exists such $\lambda(x)$ that 
$$ 
dT=0
$$ 
(the components of the metric tensor are assumed
to be differentiable on coordinates).
As a result of integration, find function $\tilde x^0$
such that $T=d\tilde x^0.$
This function define a set of surfaces
such that  each point of the domain lies on some surface.

Consider vector field $u=(1,0,0,0)^T.$
Each vector of this field is timelike, because $\tilde g_{00}>0,$
and $\tilde g_{0i}=0,$  $i=1,2,3.$
Then it can be regarded as velocity of an observer: $dx/d\tilde x^0=u.$
Trajectories of the observers are lines $x^1={\rm const},$ $x^2={\rm const},$ $x^3={\rm const}.$
Take one of the observers as the main observer.
Coordinates $\tilde x^0,$ $x^1,$ $x^2,$ $x^3$
do not form a system of coordinates associated with this reference frame,
because $\tilde g_{00}$ may be not equal to $1$ on the worldline of the main observer.
It is easy to understand that if we take coordinate $\tau$
as a solution to the differential equation $d\tau/d\tilde x^0=(\tilde g_{00}^{(m)}(\tilde x^0))^{1/2},$
then $g_{\tau\tau}^{(m)}=1,$ and
system of coordinates $\tau,x^1,x^2,x^3$ is associated with constructed reference frame.
Here, upper index $m$ means that component $g_{00}$ is taken along the worldline of the main observer.

Thus, if the metric tensor is set in some domain $D$
which satisfy the conditions of the Poincare lemma,
$g_{00}> 0,$ 
and there exists such $\lambda(x)$
that differential of the form (\ref{eq:TForm})
is equal to zero in $D,$
then one can construct
a reference frame in $D.$

\section{Accelerated and Rotating Reference Frames}
 
Consider two examples of reference frames.
As a first example,
take the reference frame 
moving with arbitrary acceleration along the axis $x$
of some Lorentzian frame
(here and further $x$ denotes one of the Cartesian coordinates
$x, y, z$).
That means that the main observer 
with which the accelerated frame is associated
moves with an acceleration
in configuration space associated with the initial reference frame.
Besides, all local observers of the accelerated frame
move with the same velocity as the main observer.
Denote by $v$ $x-$component of this velocity.

Let $v=0$ at an initial moment $t=0.$
Take the Cartesian coordinates on the temporal layer
of accelerated frame corresponding to the initial moment
as the spatial coordinates  $\tilde x, \tilde y, \tilde z$ in accelerated frame.
It is evident that $\tilde y=y,$ $\tilde z=z,$
because of the motion is directed along $x$ axis.
%(tilde indicates the accelerated frame).
Therefore, let's consider two-dimensional problem instead of four-dimensional one
($y=0,$ $z=0$).

Let 
main observer is at the origin of the initial  system of coordinates
at $t=0.$ 
Then
$$
x_0=\int\limits_0^{t_0}\beta(t)c\,dt.
$$
Here $t_0$ and $x_0$ are the time and the $x-$coordinate
of some point on the worldline of main observer
in initial system of coordinates, $\beta(t)=v/c.$
The time and the $x-$coordinate of that event 
in the system of coordinates associated with the accelerated frame are
$$
\tilde t_0=\frac sc=\frac 1c\int\limits_0^{t_0}\sqrt{g_{ik}u^iu^k}dt=\int\limits_0^{t_0}\sqrt{1-\beta(t)^2}dt
\equiv F(t_0),
\qquad \tilde x_0=0.
$$
%(tilde denotes that the values are taken for the accelerated frame).

According to the definition of the reference frame,
events which are simultaneous with $(\tilde t_0,\tilde x_0)$
in accelerated frame are also simultaneous with $(\tilde t_0,\tilde x_0)$
in the Lorentzian frame moving with the same velocity as the accelerated frame.
That means that they lie on the line
\equation
c(t-t_0)-\beta(x-x_0)=0.
\label{eq:line}
\endequation
Time for events on this line is the time of the main observer
$\tilde t=\tilde t_0$
(tilde denotes that the value is taken for the accelerated frame).
%Each point of this line has time $\tilde t=\tilde t_0.$
As distances between the local observers
of the accelerated frame in the initial frame are conserved,
the Cartesian coordinate $\tilde x_0$ is equal 
to the spacelike interval along line (\ref{eq:line})
from the point $(\tilde t_0,\tilde x_0)$
to the point $(\tilde t,\tilde x):$
$$
\tilde x^2=(x-x_0)^2-c^2(t-t_0)^2=(x-x_0)^2-\beta^2(x-x_0)^2=\gamma^{-2}(x-x_0)^2
$$
where $\gamma=(1-\beta^2)^{-1/2}.$
Thus, for a point  $(t,x)$  on line (\ref{eq:line}) we have 
\equation
t=t_0+\frac{\beta\gamma}c\tilde x,\qquad x=x_0+\gamma\tilde x
\label{eq:acceltrans}
\endequation
where $t_0$ and $x_0$ are time and coordinate of the event
which is simultaneous with the event $(\tilde t,\, \tilde x),$
and has spatial coordinate equal to $0$
in accelerated frame.

Expressing $t_0$ and $x_0$ through $\tilde t_0,$
we have
$$
t_0=F^{-1}(\tilde t_0),\quad x_0=\int\limits_0^{F^{-1}(\tilde t_0)}\beta(t)cdt.
$$
Using  expression (\ref{eq:acceltrans}),
one can calculate the components of the metric tensor
in the accelerated frame.
Differentiating  (\ref{eq:acceltrans}), we obtain
$$
\frac{\partial t}{\partial\tilde t}=\gamma(1+\gamma^3\beta'\frac{\tilde x}c),\quad
\frac{\partial t}{\partial\tilde x}=\frac{\beta\gamma}c,\quad
\frac{\partial x}{\partial\tilde t}=\beta\gamma(c+\tilde x\beta'\gamma^3),\quad
\frac{\partial x}{\partial\tilde x}=\gamma.
$$
Here the stroke means differentiating of a function on its argument.

In initial Lorentzian frame the metric tensor has components
(\ref{eq:psevdodiag}).
Applying covariant law of transformation,
we get components of the metric tensor in the accelerated frame.
Nondiagonal component is $\tilde g_{01}=0,$
as it should be in every reference frame.
From formal point of view,
this result is a consequence of the fact
that the events which are simultaneous in the accelerated frame
are also simultaneous in the comoving Lorentzian frame,
and in a Lorentzian frame the equality $\tilde g_{01}=0$ holds.
Diagonal components of the metric tensor are
$$
\tilde g_{00}=(1+\gamma^3\beta'\frac{\tilde x}{c})^2
%\approx 1+2\gamma^3\beta'\frac{\tilde x}{c}
,\qquad
\tilde g_{11}=-1.
$$

The dependence of $\tilde g_{00}$ from the spatial coordinate
results in acceleration of an isolated body in the accelerated frame,
because the dynamics equations contain derivatives of the metric tensor.
But the acceleration  will be  noticeable only when the body passes some distance.
Therefore, the accelerated frame can be regarded as locally Lorentzian frame
in some sufficiently small domain.

As the second example, consider the rotating reference frame.
Here we shall show that such frame can be regarded 
only in sufficiently small domain
such that velocities of rotation for all points of that domain are nonrelativistic.

We follow the traditional interpretation of the rotating frame
according to which cylindrical coordinates are
$\tilde r=r,$ $\tilde \varphi=\varphi+\Omega t,$ $\tilde z=z$
where $r,$ $\varphi,$ $z,$ $t$ are
the cylindrical coordinates and the time in the initial Lorentzian frame,
$\Omega$ is angular velocity of the rotation,
and the temporal coordinate is the same as in initial Lorentzian frame \cite{landau}.
As for a local observer of the rotating frame
coordinates don't change,
%$\tilde\varphi=\varphi_0,$
trajectories of their movement in the initial Lorentzian frame are
$r(t)={\rm}const,$ $\varphi(t)=\varphi_0-
\Omega t,$ $z(t)={\rm}const.$
It is obvious, that the events lying on the surfaces $t={\rm}const,$
which are simultaneous from point of view of the main observer,
are not simultaneous from the point of view of an observer
moving along one of these lines with the relativistic velocity.

Therefore the size of domain where one can use the rotating reference frame
is determined by such values of coordinate $r$ 
at which the motion along the coordinate lines of the temporal coordinate
is nonrelativistic: $|\Omega|r\ll c.$
Classical theory doesn't contain such restriction,
because surfaces $t={\rm}const$ represent surfaces of
simultaneous events from point of view of each observer.

On the other hand,
the introduced coordinates can be used at the greater values of $r,$
right up till $c/\Omega,$
as it is pointed out in Ref. \cite{landau}.
But in that case, these coordinates cannot be regarded as temporal one and 
three spatial ones. That is, this system of coordinates
is not associated with any reference frame.
It  can be also seen while calculating component $g_{0\varphi},$
which is not equal to $0.$

\section{Conclusion}

The definition of the reference frame is given.
This definition is based on such fundamental concept as the spacetime,
and is applicable in classical and relativistic cases.
%Proposed definition allows using
%wide classes of reference frames
%without restriction to inertial, uniformly accelerated or rotating frames.
The notions of the configuration space and of the system
of coordinates associated with a reference frame are introduced.
The conception of temporal and spatial coordinates is analyzed,
and it is shown that one of coordinates of the system of coordinates
associated with a reference frame 
is temporal one and the others are spatial ones.

It is demonstrated how the relativistic metric tensor can be constructed
on the base of measurements of time and distances.
It is found that 
$g_{0i}=0,$ $i=1,2,3$
in coordinates which are associated with the reference frame.
These conditions can be used as additional boundary conditions
in boundary problems  for the gravitational field equations.
The necessity of additional conditions was shown by Einstein \cite{ein2}.

Construction of a reference frame is possible
on the base of the assumption
that there are surfaces of simultaneous events
from point of view of a certain observer.
This assumption is valid for many known solutions to the gravitational field equations \cite{schmutzer},
because of
%corresponding components of the metric tensor are equal to $0:$
$g_{0i}=0,$ $i=\overline{1,3}$  for these solutions.
In the general case, as it is shown at the present work,
the existence of such surfaces is equivalent to the assertion 
that for some $\lambda(x)$ differential of the form $T=\lambda(x)(g_{00}, g_{01}, g_{02}, g_{03})$
is equal to $0.$

If the assumption being discussed is realized,
components of the metric tensor can be made to have physical sense.
The possibility of comprehension of solutions
for which spaces of simultaneous events do not exist
is questionable.
From this point of view, it is quite reasonable
to investigate only solutions with zero components $g_{0i}.$

When constructing a reference frame
in some domain of the spacetime,
we represent this domain as union
of mutually nonintersecting layers.
Such structure is well known in relativity theory as foliation.
In the works of Arnowitt, Deser, Misner \cite{adm}, and Dirac \cite{dirac},
the foliation of the spacetime was used for Hamiltonian formulation of the theory of relativity.
Besides, the foliation is widely used 
for numerical solution of the gravitational field equations (see, for example, \cite{gour}).
In \cite{gour}, components $g_{0i}$ are not restricted.
So, foliations considered in \cite{gour} have layers which are some timelike surfaces,
but not surfaces of simultaneous events.

Among various kinds of foliations, mention the geodesic foliation
when the metric tensor has components $g_{00}=1,$ $g_{0i}=0,$ $i=\overline{1,3}.$
Corresponding coordinates are called semigeodesic or Gauss normal coordinates \cite{weinberg}.
According to the approach presented in this article,
such coordinates do not form a system of coordinates
associated with any reference frame,
as $x^0$ is not temporal coordinate.
%Semigeodesic coordinates can be regarded as some analogue 
%of a system of coordinates
%associated with reference frame.
As distinct from semigeodesic coordinates,
in coordinates
associated with a reference frame,
$g_{00}$ is not equal always to $1,$ but can vary.

In the definition of the reference frame,
concepts of the main observer and of the congruence of the local observers are used.
In contrast to \cite{math,vlad}
where such concepts are also used,
additional condition is formulated.
This condition is simultaneity of all events on each temporal layer
from point of view of all observers.
It is always satisfied in the classical theory,
and results in some restrictions in relativity theory.
For example, it is shown that in relativity theory
the rotating reference frame can be used only 
for events sufficiently close to the rotation axis.
%In the works \cite{math}, \cite{vlad},
%the worldlines of the local observers are considered
%in coordinates, which are not necessarily temporal and spatial ones.
%The way of construction of such coordinates is not considered there.
%Add also, that the presented here approach is equally right
%for classical theory and for relativity theory,
%while in the cited works only relativity theory is regarded.

Note also that in various works
the concepts of the temporal coordinate and the spatial ones
are insufficiently formalized.
For example, in \cite{fok} the question was raised
when one coordinate is temporal one and others are spatial ones.
Given there conditions for components of the metric tensor
mean, in fact, that the basis vector of one of coordinates is timelike,
and the basis vectors of other coordinates are spacelike.
%in the sense that the displacements along these vectors
%give timelike and spacelike intervals correspondingly.
According to the presented here approach,
it does not mean that first coordinate is temporal one
and the others are spatial ones.
The properties of the timelikeness and the spacelikeness are absolute
and does not depend of the system of coordinates.
But a vector can be the basis vector 
for a temporal coordinate or a spatial coordinate
in some reference frame, whereas
in another reference frame it can be the basis vector for coordinate,
which is combination of temporal and spatial coordinates.
At that in \cite{fok}, the components $g_{0i},$ $i=1,2,3$  are not restricted.

\end{document}